\documentclass[12pt,preprint]{aastex}
\title{Stellar Sources for Heavy $r$-Process Nuclei}
\author{Y.-Z. Qian\altaffilmark{1} and G. J. Wasserburg\altaffilmark{2}}
\altaffiltext{1}{School of Physics and Astronomy, University of Minnesota,
Minneapolis, MN 55455; qian@physics.umn.edu.}
\altaffiltext{2}{The Lunatic Asylum, Division of Geological and Planetary
Sciences, California Institute of Technology, Pasadena, CA 91125;
isotopes@gps.caltech.edu.}

\begin{document}

\begin{abstract}
The stellar sites and the complete mechanism of $r$-process 
nucleosynthesis are still unresolved issues.
From consideration of the observed abundances in metal-poor stars, 
it is proposed that the production of the heavy $r$-process nuclei
($r$-nuclei with mass numbers $A>130$) is not related to
the production of the Fe group elements or of the elements with
lower atomic numbers: Na, Mg, Al, Si, Ca, Sc, and Ti. This requires
that the production of the heavy $r$-nuclei not occur in supernovae
with extended shell structure, but be associated with either bare
neutron stars or Type II supernovae (SNe II) in the mass range 
$8\,M_\odot<M<10\,M_\odot$. From the observations of stars with 
[Fe/H]~$\sim -3$ but with high abundances of $r$-elements, it is 
clear that these $r$-process enrichments cannot represent the 
composition of the interstellar medium from which the stars were 
formed, but must represent very local contamination from binary 
companions. Further evidence for very high enrichments of $s$-process 
elements in metal-poor stars also requires binary systems for 
explanation. We propose that the accretion-induced collapse (AIC) of 
a white dwarf into a neutron star in a binary system may be associated 
with the production
of the heavy $r$-nuclei and may provide occasional coupling of
high $r$-process and high $s$-process enrichments in the envelopes
of low-mass stars with low [Fe/H]. If we assume that
the bulk of the heavy $r$-nuclei
are produced in AIC events, then these events would have produced 
$\sim 1.6\times 10^9$ neutron stars in the Galaxy. A much larger number 
of white dwarf binaries would have resulted from the evolution of
other binary systems. The AIC scenario removes the assignment in
our earlier model that SNe II provide the bulk of the heavy $r$-nuclei
and relegates $r$-process production in SNe II to the light $r$-nuclei 
with $A\lesssim 130$. This new assignment removes the requirement 
in our earlier model that most SNe II produce no Fe and gives Fe
yields that are in accord with the observed values for most SNe II.
\end{abstract}

\keywords{nuclear reactions, nucleosynthesis, abundances ---
Galaxy: evolution --- Stars: Population II}

\section{Introduction}
In this paper we try to establish the intrinsic characteristics of
the stellar sources for the heavy $r$-process nuclei ($r$-nuclei
with mass numbers $A > 130$ corresponding to the elements Ba and
above) using aspects of a phenomenological model presented earlier
(Wasserburg \& Qian 2000; Qian \& Wasserburg 2001b, 2002). The
bulk of the light $r$-nuclei (with $A\lesssim 130$) appear to have
different sources from those for the heavy $r$-nuclei 
(Wasserburg, Busso, \& Gallino 1996) and are not extensively 
discussed here. From the
observational data on metal-poor (i.e., low [Fe/H]) stars, we will
show that the heavy $r$-nuclei are produced with only some 
coproduction of the light $r$-elements such as Sr, Y, and Zr,
but without any coproduction of the elements from Na to Zn 
(including Fe). We argue that supernovae with extended shell 
structure cannot be the source for the heavy $r$-nuclei. Instead, 
these nuclei must be produced during the formation of an essentially 
bare neutron star. Such a state may result from a low-mass 
($\sim 8$--$10\,M_{\odot}$) Type II supernova (SN II) or from the 
accretion-induced collapse (AIC) of a white dwarf into a neutron star
in a binary system. The observational basis that establishes the 
distinctive characteristics of the stellar sources for the heavy 
$r$-nuclei is discussed in \S2. In \S3 we discuss the requirement of 
an essentially bare neutron star for the production of the heavy 
$r$-nuclei and the proposed role of AIC events. The special nature 
of both $r$-process and $s$-process enrichments in binary systems 
with AIC is discussed in \S4. In \S5 we explore some of the 
consequences for our earlier model and other implications that follow 
if AIC events are a source for the heavy $r$-nuclei.

\section{Observational Basis}
Elemental abundances in halo stars of low metallicities have
provided fundamental data for understanding nucleosynthesis and
chemical evolution of the Galaxy. The observations (McWilliams et
al. 1995; Ryan, Norris, \& Beers 1996; Burris et al. 2000; Westin
et al. 2000; Sneden et al. 2000; Norris, Ryan, \& Beers 2001) led
us to propose earlier that there was a major transition at
[Fe/H]~$\sim -3$ in the production of heavy elements. The Fe production 
at [Fe/H]~$<-3$ was dominated by very massive ($\gtrsim 100\,M_\odot$) 
stars (VMSs). It was proposed that 
the VMS activities ceased and major formation of normal
stars (with masses $M\sim 1$--$60\,M_\odot$) took over when a critical
metallicity corresponding to [Fe/H]~$\sim -3$ was reached in the 
interstellar medium (ISM). This is in accord with the recent study by
Bromm et al. (2001), which indicates that substantial formation of a
normal stellar population could only occur when a metallicity 
substantially above $5\times 10^{-4}$ times the solar value was achieved
in the ISM to provide sufficient cooling for the collapse and 
fragmentation of gas clouds. The subsequent Fe production at 
$-3<{\rm [Fe/H]}<-1$ was then considered to be dominated by
SNe II. The production of the heavy $r$-elements relative to Fe was 
very low prior to the achievement of [Fe/H]~$\sim -3$ in the ISM, but 
increased sharply at [Fe/H]~$\sim -3$ and subsequently
approached a well-defined trend for normal stages of Galactic
evolution (Wasserburg \& Qian 2000; Qian \& Wasserburg 2001b, 2002). 
This behavior is shown for Eu (a predominantly $r$-process 
element in the solar system) in Figure 1a where the available data 
on $\log\epsilon({\rm Eu})$ are plotted versus
[Fe/H] \footnote{Standard spectroscopic notation is used here:
$\log\epsilon({\rm E})\equiv\log({\rm E/H})+12$, where (E/H) is
the abundance ratio of element E to hydrogen in a star, and ${\rm
[Fe/H]}\equiv\log({\rm Fe/H})-\log({\rm Fe/H})_\odot$.}. It can be
seen that for [Fe/H]~$<-2.4$ there is a very wide scatter in
$\log\epsilon({\rm Eu})$ while for higher values of [Fe/H] 
$\log\epsilon({\rm Eu})$ follows [Fe/H] in a more regular trend.
This trend is indicated by the solid line segment of unit 
slope in Figure 1a and reflects a constant Eu/Fe ratio for the net 
production by all stellar sources. The results for Eu are similar to 
those for Ba as shown in Figure 1b. The transition at [Fe/H]~$\sim -3$
is prominently demonstrated for Ba by the contrast between the data at
[Fe/H]~$<-3$ and those at [Fe/H]~$>-3$. Note that this transition 
reflects the changes in the relative production rates
of the heavy $r$-elements with respect to Fe, and therefore, can be 
used to identify the nucleosynthetic contributions to the ISM from
the sources for the heavy $r$-elements relative to those from the
sources for Fe. For example, the transition for Ba corresponds to an
increase by a factor of 10 (solid curve in Fig. 1b) to 20 
(dot-dashed curve in Fig. 1b) in the relative production of Ba with
respect to Fe from [Fe/H]~$<-3$ to [Fe/H]~$>-3$ 
(Qian \& Wasserburg 2002). However, the 
transition at [Fe/H]~$\sim -3$ by itself cannot identify 
the relative populations of the different stellar types. 

To elucidate the nature of the sources for the heavy $r$-elements, 
we exhibit in Figure 2 the
available data on HD 115444, HD 122563 (Westin et al. 2000), and
CS 31082--001 (Hill et al. 2002) with [Fe/H]~$=-2.99$, $-2.74$,
and $-2.9$, respectively, which show clear evidence for varying
degrees of $r$-process enrichment within a narrow range of [Fe/H].
We note that the data (not shown) on CS 22892--052 with [Fe/H]~$=-3.1$ 
(McWillam et al. 1995; Sneden et al. 2000) are very close to those on
CS 31082--001. We also note that high Th abundances were 
found in CS 22892--052 and HD 115444 while high Th and U abundances 
were found in CS 31082--001. It can be seen from Figure 2a that the 
abundances of the elements from Na to Zn (including Fe) in the three 
stars shown there are all rather constant with a typical range of 
$\sim 0.3$ dex in $\log\epsilon$. This is in sharp contrast to the wide 
range of $\sim 2$ dex in $\log\epsilon$ for the heavy $r$-elements
(Ba and above) as shown in Figure 2b for the same stars. Figure 2b also
shows that the relative abundances of the heavy $r$-elements in each 
star can be described remarkably well by the solar $r$-process 
abundance pattern (solar $r$-pattern). Indeed, extensive studies
(e.g., Sneden et al. 1996; Johnson \& Bolte 2001) found that the heavy 
$r$-elements in a number of metal-poor 
stars often closely follow the solar $r$-pattern while having widely
varying absolute abundances.

It follows from Figure 2 that the sources 
for the heavy $r$-nuclei are not connected with the production of the 
elements from Na to Zn, which include not only the Fe group but also 
the so-called ``$\alpha$-elements'' Mg, Si, Ca, and Ti. This 
observation was discussed in our earlier study (Qian \& Wassurburg 
2002) of the nucleosynthetic yields which we attributed to VMSs and
two hypothesized types of SNe II [SNe II($H$) and SNe II($L$)].
In particular, the abundances of the elements from Na to Zn at
[Fe/H]~$\sim -3$ were considered to represent a prompt inventory of
the elements dominantly produced by VMSs. The prompt inventory of 
the light $r$-elements Sr, 
Y, and Zr is also significant (Qian \& Wasserburg 2001b). 
This can be seen from the data on
HD 115444 and HD 122563 shown in Figure 2b. These two stars have very 
different abundances of the heavy $r$-elements but essentially the 
same abundances of Sr, Y, and Zr, which are dominated by the prompt 
inventory of these elements. Compared with HD 115444 and HD 122563, 
CS 31082--001 have much higher abundances of the heavy $r$-elements 
and significantly higher abundances of Sr, Y, and Zr (see Fig. 2b). 
This indicates that the sources for the heavy $r$-nuclei also produce 
some light $r$-nuclei and such contributions overwhelm the prompt 
inventory of Sr, Y, and Zr in CS 31082--001. The data on CS 31082--001
(see Fig. 2b) and CS 22892--052 (Sneden et al. 2000) show that the 
light $r$-elements produced by the sources for the heavy $r$-nuclei
in these two stars also include Nb, Ru, Rh, Pd, Ag, and Cd. However,
the sources for the bulk of the light $r$-nuclei appear to be different
from those for the heavy $r$-nuclei as first argued by Wasserburg et al. 
(1996) based on the inventory of $^{182}$Hf and $^{129}$I in the early
solar system. This was supported by the data on CS 22892--052 (Sneden
et al. 2000), CS 31082--001 (Hill et al. 2002), BD +17$^\circ$3248
(Cowan et al. 2002), and several other metal-poor stars (Johnson \&
Bolte 2002b), which showed that the light $r$-elements such as Ag in
these stars are clearly deficient relative to the solar $r$-pattern
translated to pass through the Eu data. The production of the
light $r$-elements appears to be associated with that of Fe. This was 
studied in some detail in Qian \& Wasserburg 
(2001b) but will not be extensively discussed here.

The range in $\log\epsilon$ for Eu and Ba in many halo stars with 
[Fe/H]~$\sim -3$ shown in Figure 1 and the range in $\log\epsilon$
for the pure heavy
$r$-elements in three such stars shown in Figure 2 require special
attention. If we consider that the solar $r$-process abundances
represent the Galactic average at the time of solar system
formation (SSF), and that the absolute yield $Y_{\rm E}/{\rm H}$
of element (or nuclide) E per H atom in a standard reference mass
of ISM which will mix with 
the nucleosynthetic products is constant for every event producing
the heavy $r$-nuclei, it follows (see Appendix A) that
\begin{equation}
({\rm E/H})_{\odot,r}=fT_\odot(Y_{\rm E}/{\rm H})
=n_{\odot}(Y_{\rm E}/{\rm H}),
\label{eh}
\end{equation}
where $({\rm E/H})_{\odot,r}$ is the solar $r$-process abundance
ratio of E to hydrogen and $f$ is the frequency for replenishing
the standard reference mass of
ISM with newly-synthesized material by the events producing
the heavy $r$-nuclei. The term $n_{\odot}=fT_\odot$ is the number
of such events per standard reference mass of
ISM over the period $T_\odot\sim 10^{10}$ yr prior to
SSF. Based on the abundance of $^{182}$Hf (with a mean lifetime of
$1.3\times 10^7$ yr) in the early solar system, $f$ was estimated
to be $\sim (10^7\ {\rm yr})^{-1}$ (Wasserburg et al. 1996; Qian
\& Wasserburg 2000). This gives $n_\odot\sim 10^3$. For a
metal-poor star formed from the ISM in the early Galaxy, the
number $n$ of the events contributing to the abundance of E in the
star is
\begin{equation}
n=n_\odot\frac{({\rm E/H})}{({\rm E/H})_{\odot,r}},
\end{equation}
where (E/H) is the number abundance ratio of E to hydrogen in the star.

For the three stars with [Fe/H]~$\sim -3$ shown in Figure 2, their 
enrichments of heavy $r$-elements correspond to $n\sim 1$--50.
While the abundances corresponding to $n\sim 1$--10 for e.g., Eu,
can be reasonably attributed to the enrichment of an average ISM
at [Fe/H]~$\sim -3$ by the events producing the heavy $r$-nuclei,
it is rather problematic to consider that $n\sim 50$ such events
could have occurred in an average ISM without any other events
also occurring to change the Fe abundance. As argued
earlier by us (Qian \& Wasserburg 2001a), the high $r$-process
enrichments in stars with very low [Fe/H] such as CS 22892--052
and CS 31082--001 (Eu and Ba data shown as filled diamonds and 
filled circles, respectively, in Fig. 1) most plausibly resulted from 
surface contamination by being a low-mass binary companion to the stellar 
source for the heavy $r$-nuclei. There is some indication that
CS 22892--052 might be in a binary (Preston \& Sneden 2001).

A special case of high Eu enrichment at low [Fe/H] was discovered
recently by J. Cohen and N. Christlieb, who found 
$\log\epsilon({\rm Eu})=0.17$ corresponding to $n\sim 460$ for
HE 2148--1247 with [Fe/H]~$=-2.3$ (Cohen et al. 2003). The data 
point (open diamond in Fig. 1a) for this star lies far above the 
trend for the evolution of Eu relative to Fe at [Fe/H]~$>-2.4$.
The result for Ba is similar (see Fig. 1b).
Upon being informed of the initial discovery and prior to the 
detection of Pb in this star, we attempted to predict the abundances 
of all the other elements from only the preliminary data on Eu and 
Fe using our earlier model (Qian \& Wasserburg 2001b, 2002), which
assumes no $s$-process contributions. These predictions are shown
along with the final data in Figure 3a. It can be seen that in
consideration of the observational errors, there is good agreement 
between the predictions and the data for most of the elements. 
This is the case especially for the elements from Mg to Ni, 
the predicted abundances of which have significant contributions 
from the prompt inventory. However, there is a serious discrepancy
for the elements Ba, La, Ce, Pr, and Nd as highlighted in Figure 3b. 
This discrepancy and the
detection of high Pb abundance in HE 2148--1247 unambiguously 
demonstrate the presence of high $s$-process enrichments in this 
star [the Pb in this star cannot be plausibly attributed to the decay
of the actinides produced by the $r$-process as the observed value
of $\log\epsilon({\rm Pb})=2.8$ far exceeds 
$\log\epsilon({\rm Th})=-0.5$]. On the other hand, 
the Ba/Eu ratio in this star is $\sim 6$
times smaller than that for the main $s$-component of the solar
abundances (Arlandini et al. 1999). We cannot construct an $s$-process
scenario that could decrease the Ba/Eu ratio greatly below the value
for the solar main $s$-component. This and the tentatively
inferred Th abundance would imply a mixture of high $s$-process 
and high $r$-process enrichments in HE 2148--1247, the explanation
of which requires a new approach and is discussed here and in 
Cohen et al. (2003).

It was well recognized that the high $s$-process enrichments 
in metal-poor stars are associated with mass transfer 
from their previous asymptotic giant branch (AGB) companions in 
binaries. Four such stars 
with confirmed binary membership are LP 625--44 with [Fe/H]~$-2.7$ 
(Aoki et al. 2000), CS 22948--027 with [Fe/H]~$=-2.5$ 
(Hill et al. 2000; Preston \& Sneden 2001), CS 22942--019 
with [Fe/H]~$=-2.7$ (Preston \& Sneden 2001), and HE 0024--2523
with [Fe/H]~$=-2.7$ (Lucatello et al. 2003). There is also
evidence for HE 2148--1247 being in a binary (Cohen et al. 2003).

From the above discussion, it is apparent that: (1) the production
of the heavy $r$-nuclei is decoupled from the production of the
elements from Na to Zn (including the ``$\alpha$-elements'' and the Fe
group); (2) surface contamination of a low-mass binary companion by
the stellar source for the heavy $r$-nuclei is a relatively common
occurrence; (3) surface contamination with $s$-process products is 
a fairly common occurrence; and (4) there seems to be a coupling of 
high $s$-process and high $r$-process enrichments in at least one 
metal-poor star.

\section{Stellar Sources for Heavy $r$-Nuclei}
The fundamental criterion for an astrophysical environment to be
the site for $r$-process nucleosynthesis is the capability of
providing a large neutron abundance on a short timescale. Two
major candidate environments are the neutrino-driven wind from
new-born neutron stars (e.g., Woosley \& Baron 1992) and the
ejecta from neutron star mergers (e.g., Freiburghaus, Rosswog,
\& Thielemann 1999). In the neutron star merger model, the material
undergoing nucleosynthesis is ejected from a bare old neutron star
that is disrupted during the merging with another neutron star or
a black hole.
Regardless of the possibility of an $r$-process, production of
the elements from Na to Zn would not be expected to occur in this
material. Thus, if neutron star mergers are a source for the heavy
$r$-nuclei, the observational requirement that the production of
these nuclei not be related to the production of the elements from
Na to Zn is satisfied. However, due to the rarity of neutron star 
mergers, very large yields per event are required to account for 
the total $r$-process abundances in the present Galaxy if these 
events are the major source for the $r$-nuclei. The required large 
yields appear to be in conflict with the level of $r$-process 
enrichment observed in metal-poor stars (Qian 2000). While the 
possibility of neutron star mergers being the $r$-process site
deserves further investigation, here we focus on the neutrino-driven 
wind from new-born neutron stars as the likely source for the heavy 
$r$-nuclei.

A neutron star can be formed from the collapse of (1) an Fe core
of stars with $M>10\,M_\odot$, (2) an O-Ne-Mg core of stars
with $M\sim 8$--$10\,M_\odot$, and (3) a white dwarf which has
been accreting material from its low-mass companion in a binary
system. The surface layers of a new-born neutron star are heated
by an enormous flux of neutrinos through the reactions $\nu_e+n\to
p+e^-$ and $\bar\nu_e+p\to n+e^+$, which results in a
neutrino-driven wind (Duncan, Shapiro, \& Wasserman 1986). This
wind was studied as a possible $r$-process site but with very
uncertain results (e.g., Woosley \& Baron 1992; Meyer et al. 1992;
Takahashi, Witti, \& Janka 1994; Woosley et al. 1994). 
Independent of these
theoretical uncertainties, the observational requirement that the
production of the heavy $r$-nuclei be decoupled from the
production of the elements from Na to Zn (including the
``$\alpha$-elements'' and the Fe group) strongly constrains the
possible association of the above neutron star formation scenarios
with the sources for the heavy $r$-nuclei. Stars with
$M>10\,M_\odot$ possess extensive shells surrounding the Fe core.
These shells contain the elements from Na to the Fe group when
they are ejected in the SN II explosion of the star subsequent to
the collapse of the Fe core (e.g., Woosley \& Weaver 1995;
Thielemann, Nomoto, \& Hashimoto 1996).
Therefore, SNe II from stars with $M>10\,M_\odot$ cannot be
associated with the source for the heavy $r$-nuclei that were
observed in the stars shown in Figure 2. Stars with $M\sim
8$--$10\,M_\odot$ have insignificant shells surrounding the
O-Ne-Mg core (Nomoto 1984). The SN II explosion resulting from the 
core collapse of such stars produces essentially no elements from 
Na to Zn (Nomoto 1987), and
therefore, can be associated with the source for the heavy
$r$-nuclei as argued earlier by us (Qian \& Wasserburg 2002; see
also Wheeler, Cowan, \& Hillebrandt 1998). The accretion-induced
collapse (AIC) of a white dwarf produces a bare neutron star which
ejects material only through the neutrino-driven wind. Thus, the
AIC events can also be a source for the heavy $r$-nuclei.

The AIC of a white dwarf was initially proposed to explain the
origin of neutron stars with low-mass binary companions (see
Canal, Isern, \& Labay 1990 for a review). The white dwarf
in an AIC event may be of C-O composition for a progenitor with
$1\,M_\odot<M<8\,M_\odot$ or of O-Ne-Mg composition for a
progenitor with $M\sim 8$--$12\,M_\odot$. [Due to tidal mass
loss, the progenitor mass range for the formation of an O-Ne-Mg
core in binary systems is wider than the range
$M\sim 8$--$10\,M_\odot$ for single stars
(Nomoto \& Kondo 1991).] Nomoto \& Kondo (1991)
showed that in both cases a broad range of mass accretion rates
would lead to AIC. Studies of nucleosynthesis in the
neutrino-driven wind associated with AIC events (Woosley \&
Baron 1992; Fryer et al. 1999) found no $r$-processing. Instead,
these studies showed that elements such as Sr, Y, and Zr are 
produced profusely by AIC events. This was used to argue that 
these events should be very rare. Consequently, AIC events have not 
been considered as a serious candidate for the $r$-process site. 
However, the conditions in the
neutrino-driven wind are sensitive to many uncertainties in the
neutrino and neutron star physics (e.g., Qian \& Woosley 1996).
In the arguments presented below, we will assume that the conditions 
in the wind associated with
AIC events are suitable for producing the heavy $r$-nuclei. We
hope that the observational evidence discussed here would stimulate
and possibly guide further theoretical studies of these events in
connection with $r$-process nucleosynthesis.

\section{AIC Events and Chemical Enrichments in Binary Systems}
As argued above, if the AIC events are a source for the heavy
$r$-nuclei, then the production of these nuclei is automatically
decoupled from the production of the elements from Na to Zn. In
addition, as these events occur in binary systems, the ejecta from
such events would naturally lead to contamination of the low-mass
companion's surface with heavy $r$-elements. As material is
ejected through the neutrino-driven wind with only a small total
mass loss in the AIC events, ablation of the companion may be
insignificant compared with the case of SNe II. Further, the AIC
model for the production of the heavy $r$-nuclei requires that
there be a coupling of $r$-process and $s$-process enrichments in
binary systems.

In general, a star evolves through the AGB phase prior to
becoming a white dwarf. This evolution results in production of
neutrons, the capture of which by the Fe seeds gives rise to the
$s$-process. 
Depending on the distribution of neutron fluence or the ratio
of neutrons to the Fe seeds, low-metallicity AGB stars may 
predominantly produce Pb (e.g., Van Eck et al. 2001) as
predicted by Gallino et al. (1998) or produce an $s$-process
pattern similar to the main $s$-component of the solar abundances
(e.g., Aoki et al. 2000), or produce some superposition of high Pb 
enrichment and the main solar $s$-component
(e.g., Johnson \& Bolte 2002a; Cohen et al. 2003). 
In any case, the $s$-process products
along with the other elements in the envelope of the AGB star may
be dumped onto the surface of a low-mass companion via mass
transfer in a binary. If the white dwarf left behind by the AGB
star subsequently accretes mass from the companion and undergoes
AIC, the surface of the companion will be enriched again, but this
time with heavy $r$-elements. It is also plausible that the ratio 
of neutrons to the Fe seeds in some AGB stars would be too low for 
any significant $s$-processing to occur\footnote{In fact, the more
massive star of a close binary system may not go through the AGB
phase (e.g., Iben 1985), and therefore, no $s$-processing would occur 
in the star. However, this star still produces a white dwarf, the AIC
of which can then provide pure $r$-process enrichments to the
low-mass companion.} (the problem of $s$-processing
at low metallicities is now attracting considerable attention).
Thus, the low-mass star 
in a binary may have a surface composition ranging from predominantly
$r$-process [e.g., possibly CS 22892--052 (Sneden et al. 2000;
Preston \& Sneden 2001)] to predominantly $s$-process [e.g., LP
625--44 (Aoki et al. 2000)] in origin. The elements C, N, and O
will then reflect the normal stellar processing in the low-mass star
(e.g., Fujimoto, Ikeda, \& Iben 2000)
and additions from the standard evolution of the originally more
massive AGB companion. It may be rather common to find metal-poor
stars with high surface enrichments of both $s$-process and
$r$-process elements. We note that a number of metal-poor stars
were observed with high $s$-process enrichments (e.g., Hill et al.
2000; Aoki et al. 2000; Johnson \& Bolte 2002a; Lucatello et al. 2003). 
Unfortunately, the
element Th, which is unique to the $r$-process, is either not
targeted for observations or is difficult to detect for these
stars. We know of only one star, HE 2148--1247, the surface
composition of which reflects a superposition of $s$-process and
$r$-process products possibly including Th (Cohen et al. 2003).

Without Th detection, significant $r$-process enrichments may
still be identified based on the observed abundance pattern of the
elements Ba and above. This is because for these elements, the
patterns produced by the $s$-process are drastically different from 
the pattern produced by the
$r$-process. A convenient measure to characterize this difference
is the Ba/Eu ratio. As mentioned earlier, observations showed that
the abundances of the heavy $r$-elements in a number of metal-poor
stars closely follow the solar $r$-pattern. This suggests that the
Ba/Eu ratio for the events producing the heavy $r$-nuclei is close
to the solar $r$-process value (Ba/Eu)$_{\odot,r}\approx 9$
(Arlandini et al. 1999). By contrast, the Ba/Eu ratio for the
solar main $s$-component is (Ba/Eu)$_{\odot,s}\approx 650$
(Arlandini et al. 1999). Due to the variations in the ratio of
neutrons to the Fe seeds, the $s$-process in low-metallicity AGB
stars will in general produce different Ba/Eu ratios from
(Ba/Eu)$_{\odot,s}$. The metallicity dependence of the $s$-process
production pattern was demonstrated by the data (Van Eck et al.
2001) on three metal-poor stars, the $s$-process abundances of
which do not follow the solar main $s$-component but are
characterized by predominant Pb enrichments as predicted by
Gallino et al. (1998). Models that were in good agreement with the
data on Zr, La, Ce, Pr, Nd, Sm, and Pb in these stars gave the
$s$-process ratios (Ba/Eu)$_s\approx 300$ for [Fe/H]~$\approx
-1.7$ and (Ba/Eu)$_s\approx 170$ for [Fe/H]~$\approx -2.5$ (Van
Eck et al. 2001). Thus, it appears that the $s$-process always
produces (Ba/Eu)$_s\gg{\rm (Ba/Eu)}_{\odot,r}$.

The Ba/Eu ratio in a star is given by
\begin{equation}
\left({{\rm Ba}\over{\rm Eu}}\right)={{\rm (Ba/Eu)}_s\over
1-\beta_r({\rm Ba})+[{\rm (Ba/Eu)}_s/{\rm (Ba/Eu)}_{\odot,r}]
\beta_r({\rm Ba})},
\label{rba}
\end{equation}
where $\beta_r({\rm Ba})$ is the fraction of Ba contributed by the
$r$-process. The Ba/Eu ratio in the star can also be given by
\begin{equation}
{\rm (Ba/Eu)}={\rm (Ba/Eu)}_s\beta_s({\rm Eu})
+{\rm (Ba/Eu)}_{\odot,r}[1-\beta_s({\rm Eu})],
\label{seu}
\end{equation}
where $\beta_s({\rm Eu})$ is the fraction of Eu contributed by the
$s$-process. A general question is when the $r$-process and the 
$s$-process contributions can be identified for a star with a
mixture of such contributions. From Equation (\ref{rba}),
the $r$-process contributions to the star can be well recognized by 
the abundance pattern for 
$\beta_r({\rm Ba})\gg{\rm (Ba/Eu)}_{\odot,r}/{\rm(Ba/Eu)}_s$. 
From Equation (\ref{seu}),
the $s$-process contributions to the star can be identified for
$\beta_s({\rm Eu})\gg{\rm (Ba/Eu)}_{\odot,r}/{\rm (Ba/Eu)}_s$.
This is illustrated in Figure 4 by considering mixtures of the
solar $r$-pattern and the solar main $s$-component. For such mixtures,
the reference value to be compared with $\beta_r({\rm Ba})$ and
$\beta_s({\rm Eu})$ is 
${\rm (Ba/Eu)}_{\odot,r}/{\rm (Ba/Eu)}_{\odot,s}\approx 0.014$.
The dot-dashed curve in Figure 4a represents a mixture in which the 
Ba receives
equal contributions from the $r$-process and the $s$-process
[$\beta_r({\rm Ba})=0.5]$. It can be seen that this curve closely
follows the solar $r$-pattern (thick solid curve) and the $s$-process
contributions are essentially obscured. This can be understood as only 
$\approx 1.4\%$ of the Eu [$\beta_s({\rm Eu})\approx 0.014]$ is contributed 
by the $s$-process. It follows that if $\gtrsim 50\%$ of the Ba in a 
star is from the $r$-process, then the abundances of all the elements 
above Ba will exhibit the solar $r$-pattern. The case of 
$\beta_r({\rm Ba})=0.5$ corresponding to 
$\beta_s({\rm Eu})\approx 0.014$ is also shown in Figure 4b. 
For comparison, the thin solid curve in Figure 4b
represents a mixture in which the Eu receives equal contributions
from the $r$-process and the $s$-process [$\beta_s({\rm Eu})=0.5$]. 
It can be seen that this curve closely follows the solar main
$s$-component (dashed curve). We emphasize that
if $\gtrsim 50\%$ of the Eu in a star is from the $s$-process, the 
overall abundance pattern will not be easily distinguishable from an 
$s$-pattern. The identification of the $r$-process and the
$s$-process contributions is also complicated by the observational errors 
($\sim \pm 0.1$ dex in $\log\epsilon$) and by the possible variations
of (Ba/Eu)$_s$. In very uncertain cases, the $r$-process contributions 
can only be identified by the data on Ir, Th, and U, which receive much 
less or no contribution from the $s$-process compared with Eu.

The observed abundance pattern of the elements from Ba to Dy except for
Gd in HE 2148--1247
can be accounted for by a mixture of the solar $r$-pattern and the solar
main $s$-component with $\beta_s({\rm Eu})\approx 0.14$ corresponding to
$\beta_r({\rm Ba})=0.077$. This mixture is shown as the dot-dashed 
curve in Figure 3b. We cannot
offer an explanation for the offset of the Gd data from the dot-dashed
curve. For the mixture represented by this curve, 
both $\beta_r({\rm Ba})$ and 
$\beta_s({\rm Eu})$ greatly exceed the reference value of
${\rm (Ba/Eu)}_{\odot,r}/{\rm (Ba/Eu)}_{\odot,s}\approx 0.014$.
Consequently, the presence of both the $r$-process
and the $s$-process contributions in HE 2148--1247
can be established based on the observed Ba/Eu ratio alone. 

As the dominant $s$-process sources in the Galaxy are long-lived
low-mass stars, the $s$-process contributions to the ISM are
insignificant at low metallicities corresponding to very early
times. As mentioned earlier, it was well recognized that the high 
enrichments of
$s$-process elements in metal-poor stars are associated with mass
transfer from their previous AGB companions in binaries. It was
thought that all the stars with high $s$-process enrichments
should have a white dwarf companion now. Some such stars indeed
show shifts in their radial velocity (Aoki et al. 2000; Preston \&
Sneden 2001; Lucatello et al. 2003). However, 
Preston \& Sneden (2001) showed that these
cases are not common and a number of highly $s$-process enriched
stars appear to be single stars. According to the above discussion
of the AIC events, the highly $s$-process enriched metal-poor
stars, if in binaries, may have neutron stars instead of white
dwarfs as companions. Further, the neutron star formed from the
AIC of a white dwarf may receive a large kick from, e.g.,
asymmetric neutrino emission (e.g., Lai \& Qian 1998). This would
disrupt the binary and may explain the observed single stars with
high $s$-process enrichments. The binary disruption scenario would 
also imply the existence of highly $r$-process enriched single stars 
as significant $s$-processing might not occur in some AGB companions
(see footnote 4).

\section{Discussion and Conclusions}
The data on metal-poor stars require that the production of the
heavy $r$-nuclei be decoupled from the production of the elements
from Na to Zn (including the ``$\alpha$-elements'' and the Fe group).
This appears to exclude SNe II with progenitors of
$M>10\,M_\odot$, or more generally any SNe with extended shell
structure, from being the source for the heavy $r$-nuclei.
Considering that the heavy $r$-nuclei are produced in the
neutrino-driven wind from a new-born neutron star, we have argued
that only SNe II from stars with $M\sim 8$--$10\,M_\odot$ and AIC
events in binary systems, which produce essentially bare neutron
stars, satisfy the observational requirement. While neutron star 
mergers may also satisfy this requirement, they are not considered
here as a major source for the heavy $r$-nuclei. This is because 
the rarity of neutron star mergers requires very large yields per 
event, which appear to be in conflict with the observed
level of $r$-process enrichments in metal-poor stars. Further,
neutron star mergers cannot produce an intrinsic coupling between
$r$-process and $s$-process enrichments in binary systems.
As argued above, the AIC events may
naturally lead to contamination of the low-mass binary companions'
surface with both $s$-process and $r$-process products. The
$s$-process contamination is associated with the AGB progenitor
for the white dwarf which subsequently undergoes AIC into a
neutron star. Depending on the distribution of the ratio of neutrons 
to the Fe seeds that results from the evolution of
the AGB progenitor, the surface composition of the low-mass
binary companion can range from predominantly $r$-process to
predominantly $s$-process in origin. The discovery of a
highly-enriched mixture of $s$-process and $r$-process elements
including possibly Th in HE 2148--1247 provides strong support for
the AIC model for the production of the heavy $r$-nuclei. In general,
as the binary might be disrupted during an AIC event, this might
explain the existence
of single stars with high $s$-process enrichments. The binary 
disruption scenario would also 
imply the existence of highly $r$-process enriched single stars as 
significant $s$-processing might not occur in some AGB companions
(see footnote 4).
Subsequent to the disruption of the binary, the highly $s$-process 
or $r$-process enriched stars may acquire rather large proper motion.

The AIC model can be tested by discovering (1) more metal-poor
stars with high enrichments of both $s$-process and $r$-process
elements including Ir, Th, and U, (2) neutron star companions of such
stars, and (3) large proper motion of such stars in single
configuration. As some light $r$-elements are produced along with
the heavy ones in lesser quantities (Qian \& Wasserburg 2001b), 
the detection of the
light $r$-elements such as Ag may help the identification of 
$r$-process enrichments in a star.
The AIC events have no optical display due to
the lack of an envelope. Direct observations of these events
must rely on the detection of neutrinos from the neutron star
formed in such events. Gamma rays from decay of the progenitors
of the heavy $r$-nuclei, as well as radio signals from the
shocked ejecta and the neutron star, may also be observable from an
AIC event.

There are several questions regarding the generality of the sources
for the heavy $r$-nuclei that are proposed here based on observations
of metal-poor stars. It is not evident that all the heavy $r$-nuclei
in the Galaxy were produced by low-mass ($\sim 8$--$10\,M_\odot$)
SNe II and AIC events. The mass range of $\sim 8$--$10\,M_\odot$
corresponds to only $\sim 28\%$ of all the stars with
$M\sim 8$--$60\,M_\odot$ for a Salpeter initial mass function (IMF)
of the form $dN/dM\propto M^{-2.35}$. For the low-mass SNe II to
constitute $\sim 50\%$ of all SNe II would require a drastic change
from $-2.35$ to $-4.11$ in the exponent of the IMF. Thus, 
low-mass SNe II are perhaps not the major source for the heavy
$r$-nuclei due to the narrow mass range. If the proposed AIC model 
is not responsible for the bulk
of the heavy $r$-nuclei, either, then the
question is --- what are the major sources for these nuclei? If the
AIC events are the dominant or sole source for the heavy $r$-nuclei,
then it is necessary to provide a detailed
description of the $r$-process
production and the chemical enrichment associated with this scenario.

First consider the issue of chemical enrichment. In our earlier model 
(Wasserburg \& Qian 2000; Qian \&
Wasserburg 2001b, 2002), we attributed the heavy $r$-nuclei to
$\sim 90\%$ of SNe II [SNe II($H$)] and the bulk of the light
$r$-nuclei to the rest [SNe II($L$)]. In addition, it was required
that SNe II($L$) but not SNe II($H$) produce the elements from Na to 
Zn (including Fe). For the model proposed here 
with AIC events being the major
source for the heavy $r$-nuclei, there is no need to require that
most SNe II produce no elements from Na to Zn. Thus, all SNe II
from stars with $M>10\,M_\odot$ could be SNe II($L$). This removes
the conflict between our earlier model and observations of SN II
light curves, which showed that most SNe II produce $\sim
0.1\,M_\odot$ of Fe (e.g., Table 1 in Sollerman 2002). With this Fe
yield and a Galactic SN II rate of $\sim (30\ {\rm yr})^{-1}$, 
$\sim 3\times 10^8$ SNe II over the Galactic history of 
$\sim 10^{10}$ yr would enrich
the total baryonic mass of $\sim 10^{11}\,M_\odot$ in the Galaxy
with $\sim 1/3$ of the solar Fe mass fraction ($\approx 10^{-3}$).
This is in accord with the expected additional Fe contributions
from SNe Ia.

As in the case of SNe II, the amount of ISM which will dilute the 
nucleosynthetic products 
is determined by the total energy injected into the ISM. The energy
injected by an AIC event is perhaps comparable to that by an SN II
(Woosley \& Baron 1992), which corresponds to a standard dilution mass 
of $\sim 3\times 10^4\,M_\odot$ (e.g., Thornton et al. 1998). If
$n_\odot\sim 10^3$ AIC events provided this dilution mass with the
solar abundances of the heavy $r$-nuclei (see \S2), then a total
of $\sim 3\times 10^9$ AIC events must have occurred to provide
the same enrichment to the total baryonic mass of $\sim
10^{11}\,M_\odot$ in the Galaxy. This is $\sim 10$ times more than
the total number of SNe II over the Galactic history, and therefore,
suggests that the Galactic rate of AIC events must be
$\sim (3\ {\rm yr})^{-1}$ if these events are the major source for
the heavy $r$-nuclei. The value of $n_\odot\sim 10^3$ used to obtain
this rate corresponds to a frequency of 
$\sim (10^7\ {\rm yr})^{-1}$ for replenishment of the heavy $r$-nuclei
in the standard dilution mass. This frequency was inferred from the
original measurements of the $^{182}$W/$^{183}$W deficiencies in iron
meteorites relative to the earth's crust, which imply
($^{182}$Hf/$^{180}$Hf$)_{\rm ESS}=2.8\times 10^{-4}$ in the early 
solar system. Recent measurements by Yin et al. 
(2002) and Kleine et al. (2002) give 
($^{182}$Hf/$^{180}$Hf$)_{\rm ESS}=10^{-4}$. Repeating the argument 
of Wasserburg et al. (1996) for this new value indicates that
the frequency for replenishment of the heavy $r$-nuclei should be
reduced by a factor of $\sim 2$, and therefore,
the required Galactic rate for 
AIC events to be the major source for the heavy $r$-nuclei is 
$\sim (6\ {\rm yr})^{-1}$.

If the Galactic rate of AIC events is indeed 
$\sim (6\ {\rm yr})^{-1}$, then the neutrino signals from such an event 
will be observed in the near future by detectors such as Super 
Kamiokande (SK).
The neutrino signals from an AIC event should
be similar to those from an SN II. The dominant signals in a water
\v Cerenkov detector such as SK are caused by $\bar\nu_e+p\to n+e^+$. 
Eleven $\bar\nu_e$ events were observed from SN 1987A at a distance of
50 kpc by the Kamiokande II water detector with a fiducial mass of
2.14 kton (Hirata et al. 1987). Thus, if an AIC event occurs at a 
distance of 10 kpc,
$\sim 11(50\ {\rm kpc}/10\ {\rm kpc})^2(32\ {\rm kton}/2.14\ {\rm kton})
\sim 4000$ $\bar\nu_e$ events will be observed by SK 
with a fiducial mass of 32 kton.

The AIC events were originally proposed to explain the formation
of low-mass X-ray binaries (LMXBs), which consist of a low-mass star 
and a neutron star (see Canal et al. 1990 for a review). However, 
we cannot connect the AIC events with LMXBs based on the required
Galactic rate for these events to be the major source for the heavy 
$r$-nuclei. There are $\approx 120$ LMXBs observed in the Galaxy 
(van Paradijs 1995). If the lifetime of an LMXB is 
$\sim 10^8$--$10^9$ yr, the birth rate of LMXBs is
$\sim (10^6$--$10^7\ {\rm yr})^{-1}$
(Verbunt \& van den Heuvel 1995). This is far below the Galactic
rate of SNe II and that of AIC events discussed above.
Therefore, if the AIC events are the major source for the heavy
$r$-nuclei, these events may have little to do with the formation
of LMXBs. We note that the observed LMXBs have very short orbital
periods of $\sim 0.2$--400 hr (White, Nagase, \& Parmar 1995)
while the binary systems containing low-mass stars with high
$s$-process enrichments tend to have much longer orbital 
periods\footnote{An exception is the system containing
HE 0024--2523 with an orbital period of 3.14 days.} of $>1$ yr
(Aoki et al. 2000; Preston \& Sneden 2001; Cohen et al. 2003).
It is plausible that the AIC events frequently occur in wide
binary systems while LMXBs are the rare outcome of the evolution of
close binary systems.

The Galactic birth rate of pulsars, which are rotating neutron stars
with high magnetic fields, was estimated to be
(60--330 yr)$^{-1}$ (Lyne et al. 1998). It is conventional to consider
all pulsars as the products of SNe II. As only $\sim 10\%$ of the 
catalogued SN II remnants in the Galaxy are observed to contain pulsars
(Kaspi \& Helfand 2002), the Galactic SN II rate of
$\sim (30\ {\rm yr})^{-1}$ is only consistent with
the lower end of the estimated pulsar birth rate. If AIC events
are the major source for the heavy $r$-nuclei, the required Galactic
rate of $\sim (6\ {\rm yr})^{-1}$ for these events could easily
accommodate the higher end of the estimated pulsar birth rate.
In addition, accretion onto the white dwarf in an AIC event would
provide a natural mechanism to produce a rapidly rotating neutron star,
which is required for the pulsar mechanism.

According to the above discussion, if AIC events are the major
source for the heavy $r$-nuclei, their occurrences in the Galaxy
must have produced $\sim(10^{10}\ {\rm yr})/(6\ {\rm yr})
\sim 1.6\times 10^9$ neutron stars, which are
$\sim 5$ times more than those produced by SNe II. This implies
that AIC events are a frequent outcome of binary evolution, far
more common than SNe Ia. 
On the other hand, progenitor systems of AIC events
only constitute a fraction of all binary systems, most of which
would evolve into white dwarf binaries. Thus, the total number
of white dwarf binaries must greatly exceed $\sim 1.6\times 10^9$,
which is the number of neutron stars produced by AIC events.
These white dwarf binaries could
contribute significantly to the inventory of MACHOs.

As discussed in \S4, AIC events would lead to high 
$s$-process and high $r$-process enrichments in binary systems. 
These enrichments are not
diagnostic of the composition of the ISM from which the binary 
systems were formed.
There are three distinct times involved here, which correspond to
the formation of the binary, the development of the AGB phase, and 
the AIC event. We note that there may be significant delays between 
the first and the other two times depending on the stellar masses. 
Due to the time ($\gtrsim 10^9$ yr) required for the 
evolution of the most abundant stars ($M<2\,M_\odot$) to reach the 
AGB phase, it is plausible that there were no significant $s$-process 
contributions to the ISM at low [Fe/H].
Prior to the onset of SN Ia contributions at
[Fe/H]~$\sim -1$, the elements from Na to Zn can only be 
produced by stars with $M>10\,M_\odot$. We consider these elements
instead of the heavy $r$-elements or the $s$-process elements
as more reliable monitors of the 
ISM because surface contamination is not likely an issue in 
interpreting the stellar data on these elements. Consequently,
the abundances of the elements from Na to Zn 
can be plausibly associated with ongoing ``average'' evolution of 
the ISM at $-3<{\rm [Fe/H]}<-1$.

As we have emphasized here the role of binary systems at low [Fe/H], 
we must consider the effects of SNe Ia, which produce dominantly the
Fe group elements but not the ``$\alpha$-elements.'' The observed transition
in the trend for evolution of Mg, Si, Ca, and Ti relative to Fe at
[Fe/H]~$\sim -1$ indicates that major Fe contributions from SNe Ia 
only occur at [Fe/H]~$\gtrsim -1$. The
role of the AIC events as the major source for the heavy
$r$-nuclei especially at [Fe/H]~$<-1$ then implies severe
restrictions on binary evolution. It is possible that SNe Ia may only 
occur through the merging of two white dwarfs, which are produced in
binary systems with two low-mass stars (e.g., Iben 1985). In this case,
the onset of SN Ia contributions would require a delay of 
$\sim 10^9$ yr similar to that for the onset of major $s$-process 
contributions to the ISM. The parameter space for the mass
and the accretion rate of a C-O or O-Ne-Mg white dwarf leading to
AIC or an SN Ia was studied by Nomoto \& Kondo (1991). More
studies of this kind and of binary evolution in general are much
needed to assess the possibility of the predominance of AIC events
over SNe Ia at [Fe/H]~$<-1$.

Another crucial issue to be addressed is the $r$-process
production by AIC events. The diverse nature of the $r$-process
was inferred from the meteoritic data on $^{182}$Hf and $^{129}$I
by Wasserburg et al. (1996). This was supported by observations of
metal-poor stars, which showed that the light $r$-elements such as
Rh and Ag in CS 22892--052 (Sneden et al. 2000) and CS 31082--001
(Hill et al. 2002) are deficient relative to the solar $r$-pattern
translated to pass through the Eu data (see also Cowan et al. 2002;
Johnson \& Bolte 2002b). So far ab initio approaches are unable to
define the $r$-process yield patterns. Observations (Sneden et al.
1996; Westin et al. 2000; Johnson \& Bolte 2001) showed that the
heavy $r$-elements in a number of metal-poor stars closely follow
the solar $r$-pattern. This comparison between the pattern
resulting from a very small number of $r$-process events (only
a single event in some cases) at low [Fe/H]
and the solar $r$-pattern resulting from many events indicates
that the production of the heavy $r$-nuclei is well behaved and
not very sensitive to the individual production site. One possible
means of obtaining uniform yield patterns for the heavy $r$-nuclei
would be fission cycling, which tends to produce a robust pattern
with two peaks at $A\sim 130$ and 195, respectively (e.g.,
Freiburghaus et al. 1999). The observational data do not require
an invariant yield pattern of the heavy $r$-nuclei. For example,
the Th/Eu ratio in CS 22892--052 differs from that in CS 31082-001
by an amount that is too large to be explained by the possible age
difference between the two stars (Qian 2002). However, if Th is
produced but if fission cycling does not occur, the initial ratio of
neutrons to the seed nuclei for the $r$-process is still rather
restricted. This limits but does not fix the resulting yield 
pattern. It may be possible to obtain a regular but not fixed 
pattern for the heavy $r$-nuclei including Th without
fission cycling (e.g., Qian 2002).

The basic physics of neutron star formation and neutrino emission 
is essentially the same for both SNe II and AIC events. Neither
process is sufficiently understood to provide a reliable
prediction for $r$-process nucleosynthesis. It is hoped that a 
focused effort on AIC models might be fruitful in exploring the 
problem of $r$-process nucleosynthesis without consideration of 
the complexity of SN II dynamics. The neutrino-driven wind associated with
AIC events requires further studies. Among the important physics 
issues, effects of neutrino flavor mixing (e.g., Caldwell, Fuller,
\& Qian 2000) and neutron star magnetic 
field on the conditions in the wind deserve special attention.

In conclusion, we consider that a strong and coherent case can be
made for the production of the heavy $r$-nuclei by AIC events in
binary systems based on observations of metal-poor stars with
[Fe/H]~$\sim -3$. More specifically, the production of the heavy
$r$-nuclei occurs in the neutrino-driven wind from the neutron
star produced by the AIC of a white dwarf in a binary system.
This scenario provides a coupling between the $r$-process and the
$s$-process as the progenitor for the white dwarf may produce
$s$-process elements during the AGB phase.
Whether this scenario with AIC may also be the dominant source for the
heavy $r$-nuclei in the Galaxy remains to be investigated. With
regard to our earlier model, the assignment of heavy $r$-element
production to SNe II($H$) must be changed to the AIC events. The
requirement of a major change in the mode of star formation at
[Fe/H]~$\sim -3$ and the decoupling of the production of the
heavy $r$-nuclei from that of Fe appear to be firm. 
Prior to [Fe/H]~$\sim -3$ being obtained in the ISM, VMSs with
very large dilution masses played a dominant role in chemical
evolution. The VMS activities ceased and major formation of normal
stars took over at
[Fe/H]~$\sim -3$. The evolution of neutron-capture elements at
$-3\lesssim{\rm [Fe/H]}<-1$ must be reinvestigated now that some
coupling between the $r$-process and the $s$-process has been
established. In particular, our previous assumption of no
significant $s$-process contributions at [Fe/H]~$<-1$ and our
inferred modifications of the $s$-process and the $r$-process
components of the solar abundances must be reevaluated.

\acknowledgments         
We thank Gerry Brown for calling our attention to the important 
issue of LMXBs, Petr Vogel for discussion of supernova neutrino
detection, and an anonymous referee for helpful comments.
This work was supported in part by DOE grants DE-FG02-87ER40328 and
DE-FG02-00ER41149 (Y.Z.Q.) and by NASA grant NAG5-11725 (G.J.W.),
Caltech Division Contribution 8899(1101).

\appendix
\section{Chemical Evolution Model}
Equation (\ref{eh}) can be derived from a simple model that treats the
evolution of the abundance of element E in the ISM. Consider a parcel
of matter initially consisting of gas only. We assume that the gas of
the parcel is well mixed and its amount can only decrease at subsequent
times due to astration or escape from the parcel. At time $t$, 
the total numbers of E and hydrogen atoms in the gas of the parcel 
are (E) and (H), respectively. The evolution of (E) is governed by
\begin{equation}
{d({\rm E})\over dt}=\sum_iP_i({\rm E})+\left({{\rm E}\over{\rm H}}\right)
{d({\rm H})\over dt},
\label{pe}
\end{equation}
where $P_i({\rm E})$ is the number of E atoms injected into the gas per
unit time from prototypical sources of type $i$, and $d({\rm H})/dt<0$ 
represents the total rate of loss of gas due to astration and escape from 
the parcel. Equation (\ref{pe}) can be rewritten as
\begin{equation}
{d({\rm E/H})\over dt}=\sum_i{P_i({\rm E})\over({\rm H})}=\sum_i
{Y_i({\rm E})\over{\rm H}}{dn_i\over dt},
\label{ye}
\end{equation}
where the yield $Y_i({\rm E})/{\rm H}$ is the number of E atoms produced 
by a single event of type $i$ per H atom in a standard reference mass of
gas which will mix with the nucleosynthetic products from such an event.
This yield is assumed to be the same for all such events.
In equation (\ref{ye}), $n_i$ is the number of the events of type $i$
that have occurred in the standard reference mass of gas up to time $t$.
Equation (\ref{ye}) can be solved to give
\begin{equation}
\left({{\rm E}\over{\rm H}}\right)=\left({{\rm E}\over{\rm H}}\right)_P
+\sum_i{Y_i({\rm E})\over{\rm H}}n_i,
\label{ehi}
\end{equation}
where (E/H)$_P$ is the initial value and corresponds to the prompt inventory
of E that exists in the gas before normal astration takes place.
Equation (\ref{eh}) is obtained for a single type of source when the prompt
inventory is overwhelmed by the subsequent production.

If two parcels of matter with different evolutionary histories are mixed,
then the equation for (E/H) is of the same form as in equation (\ref{ehi})
but $n_i$ becomes the weighted value $\tilde n_i$ for the mixture. If matter
with no enrichment is mixed with an ISM that has been enriched in E, then 
both the resultant $\tilde n_i$ and (E/H) are decreased. It is not possible 
to decrease $\tilde n_i$ and keep (E/H) fixed. The abundance patterns 
resulting
from mixed or unmixed parcels is unchanged as long as the yield pattern of
each type and the ratios $n_i/n_j$ for all contributing types are fixed. 
See Qian \& Wasserburg (2001b) for a full treatment.

If we consider infall of unenriched gas during the evolution of a parcel,
then the term $d({\rm H})/dt$ in equation (\ref{pe}) is replaced by
$d({\rm H})/dt-d({\rm H})_{\rm in}/dt$ with $d({\rm H})_{\rm in}/dt$ being
the rate of infall. This gives
\begin{equation}
{d({\rm E/H})\over dt}=\sum_i{Y_i({\rm E})\over{\rm H}}{dn_i\over dt}
-\left({{\rm E}\over{\rm H}}\right){1\over({\rm H})}
{d({\rm H})_{\rm in}\over dt}.
\label{inf}
\end{equation}
The solution to equation (\ref{inf}) depends in detail on the rate of infall 
and can be formally written as
\begin{equation}
\left({{\rm E}\over{\rm H}}\right)=\left({{\rm E}\over{\rm H}}\right)_P
+\sum_i{Y_i({\rm E})\over{\rm H}}n_i-\int_0^t
\left({{\rm E}\over{\rm H}}\right){1\over({\rm H})}
{d({\rm H})_{\rm in}\over dt'}dt'.
\label{ehinf}
\end{equation}
Compared with the case without infall, equation (\ref{ehinf}) gives a lower
value of (E/H) for a given $n_i$ or requires larger values of $n_i$ for
a fixed (E/H). However, the ratio of the $n_i$ values corresponding to two
different values of (E/H) may not be significantly affected by infall.
The argument presented in the text below equation (\ref{eh}) is based on
the ratio $n/n_\odot$, and therefore, is not substantially altered. The 
basic problem discussed there has to do with the observations that different
metal-poor stars with essentially the same [Fe/H] have widely varying
inventories of heavy $r$-nuclei. In some cases, the observed abundances of
these nuclei are so high that they cannot be attributed to enrichments of
the ISM from which the stars were formed but must have resulted from surface
contamination in binary systems.

\clearpage
\figcaption{(a) Data on $\log\epsilon({\rm Eu})$ versus [Fe/H] for
metal-poor stars (filled squares: McWilliam et al. 1995; open circles:
Burris et al. 2000; open triangles: Westin et al. 2000; filled diamond:
Sneden et al. 2000; asterisks: Johnson 2002; filled circle: Hill et al.
2002; filled triangle: Cowan et al. 2002; open diamond: Cohen et al. 
2003). The dotted line indicates $\log\epsilon_\odot({\rm Eu})$ in the 
sun. The solid line segment of unit slope indicates
the mean trend for evolution of Eu relative to Fe at 
$-2.4<{\rm [Fe/H]}<-1$. A line of unit slope corresponds to additions
of Eu and Fe to the big bang debris at a constant production ratio.
Note the large scatter and the abrupt departure from the dashed extension
of the solid line segment in the region of $-3<$~[Fe/H]~$<-2.4$ between
the two vertical dashed lines. The open diamond 
represents the data on HE 2148--1247 (Cohen et al. 2003) and
clearly lies far above the trend. (b) Data on $\log\epsilon({\rm Ba})$ 
versus [Fe/H] for metal-poor stars. Data symbols are the same as in
(a) except that the filled squares now represent the data from 
McWilliam (1998) and the asterisks now represent the data from 
Ryan et al. (1996) and Norris et al. (2001). The dotted line indicates 
$\log\epsilon_\odot({\rm Ba})$ in the sun. The solid and dot-dashed
curves indicate the mean trend for evolution of Ba relative to Fe at 
$-4<{\rm [Fe/H]}<-1$. These two curves correspond to increases by 
factors of 10 (solid curve) and 20 (dot-dashed curve) in the relative 
production of Ba with respect to Fe at [Fe/H]~$=-3$. The dashed line
of unit slope indicates the mean trend for evolution of Ba relative to 
Fe at $-2.4<{\rm [Fe/H]}<-1$ and is far above the data at
$-4<$~[Fe/H]~$<-3$. Note again the large scatter in the region of 
$-3<$~[Fe/H]~$<-2.4$ between
the two vertical dashed lines. The open diamond for
the data on HE 2148--1247 (Cohen et al. 2003) clearly lies far above 
the trend.}

\figcaption{Data on the elements from C to Pt in HD 115444 (filled
circles), HD 122563 (squares: Westin et al. 2000), and CS 31082-001 
(asterisks: Hill et al. 2002) with
[Fe/H]~$=-2.99$, $-2.74$, and $-2.9$, respectively. 
(a) The $\log\epsilon$ values for the elements from C to Ge are
shown. The data on CS 31082-001 are connected with solid line
segments as a guide. Missing segments mean incomplete data. The
downward arrow at the asterisk for N indicates an upper limit.
Note that the $\log\epsilon$ values for the elements from Na to Zn
are almost indistinguishable for the three stars.
(b) The $\log\epsilon$ values for the elements from Sr to Pt are
shown. The data on CS 31082-001 to the left of the vertical
dotted line are again connected with solid line segments as a guide.
In the region to the right of the vertical dotted line, the solid,
dot-dashed, and dashed curves are the solar $r$-pattern translated 
to pass through the Eu data for CS 31082-001, HD 115444, and
HD 122563, respectively. Note the close description of the data
by these curves. The shift between the solid and the dashed
curves is $\sim 2$ dex.}

\figcaption{(a) Comparison of blind predictions (filled circles connected 
with solid curves) from the phenomenological model of Qian \& Wasserburg
(2001b, 2002) with the data (squares with error bars) on HE 2148--1247
(Cohen et al. 2003). 
The predictions were made from the preliminary values of 
$\log\epsilon({\rm Fe})=5.16$ and $\log\epsilon({\rm Eu})=0.34$, which
are close to the data on Fe and Eu from the final analyses. 
Note the close agreement between the predictions and the data
for the elements from Mg to Ni and for Sr, Y, and Zr. Also note
the large discrepancy for the elements from Ba to Nd. No predictions
were made for C, N, and Pb. The downward arrow at the data symbol for
Th indicates only a possible detection. (b) The data on the elements
from Ba to Dy are compared with the solar $r$-pattern (thin solid curve)
and the solar main $s$-component (dotted curve), both of which are
translated to pass through the Eu data. The predictions in (a) are
close to the translated solar $r$-pattern and not shown. It is clear
that the data must represent a mixture of the solar $r$-pattern and
the solar main $s$-component. The dot-dashed curve is such a mixture
with 14\% of the Eu contributed by the $s$-process 
[$\beta_s({\rm Eu})=0.14$]
and describes the data very well except for Gd.}

\figcaption{(a) Comparison of mixtures having a fixed number of Ba atoms
with different proportions from the $r$-process and the $s$-process:
pure $s$-process represented by the solar main $s$-component [dashed
curve, $\beta_r({\rm Ba})=0$], pure $r$-process represented by the solar 
$r$-pattern [thick solid curve, $\beta_r({\rm Ba})=1$], and a mixture 
with equal contributions to Ba from the $r$-process and the $s$-process
[dot-dashed curve, $\beta_r({\rm Ba})=0.5$]. Note that the 
dot-dashed curve closely follows the solar $r$-pattern. (b) Same as (a)
but the number of Eu atoms is fixed instead
with different proportions from the 
$r$-process and the $s$-process: pure $s$-process [dashed curve, 
$\beta_s({\rm Eu})=1$], pure $r$-process [thick solid curve, 
$\beta_s({\rm Eu})=0$], and a mixture with equal contributions to 
Eu from the $r$-process and the $s$-process [thin solid curve, 
$\beta_s({\rm Eu})=0.5$]. Note that the thin solid curve closely follow
the solar main $s$-component except for the element Ir. The dot-dashed 
curve for the mixture in (a) with 50\% of the Ba from the $r$-process
is also shown in (b). This mixture corresponds to only $\approx 1.4\%$
of the Eu being contributed from the $s$-process 
[$\beta_s({\rm Eu})=0.014$], and therefore, stays close to the
solar $r$-pattern.}
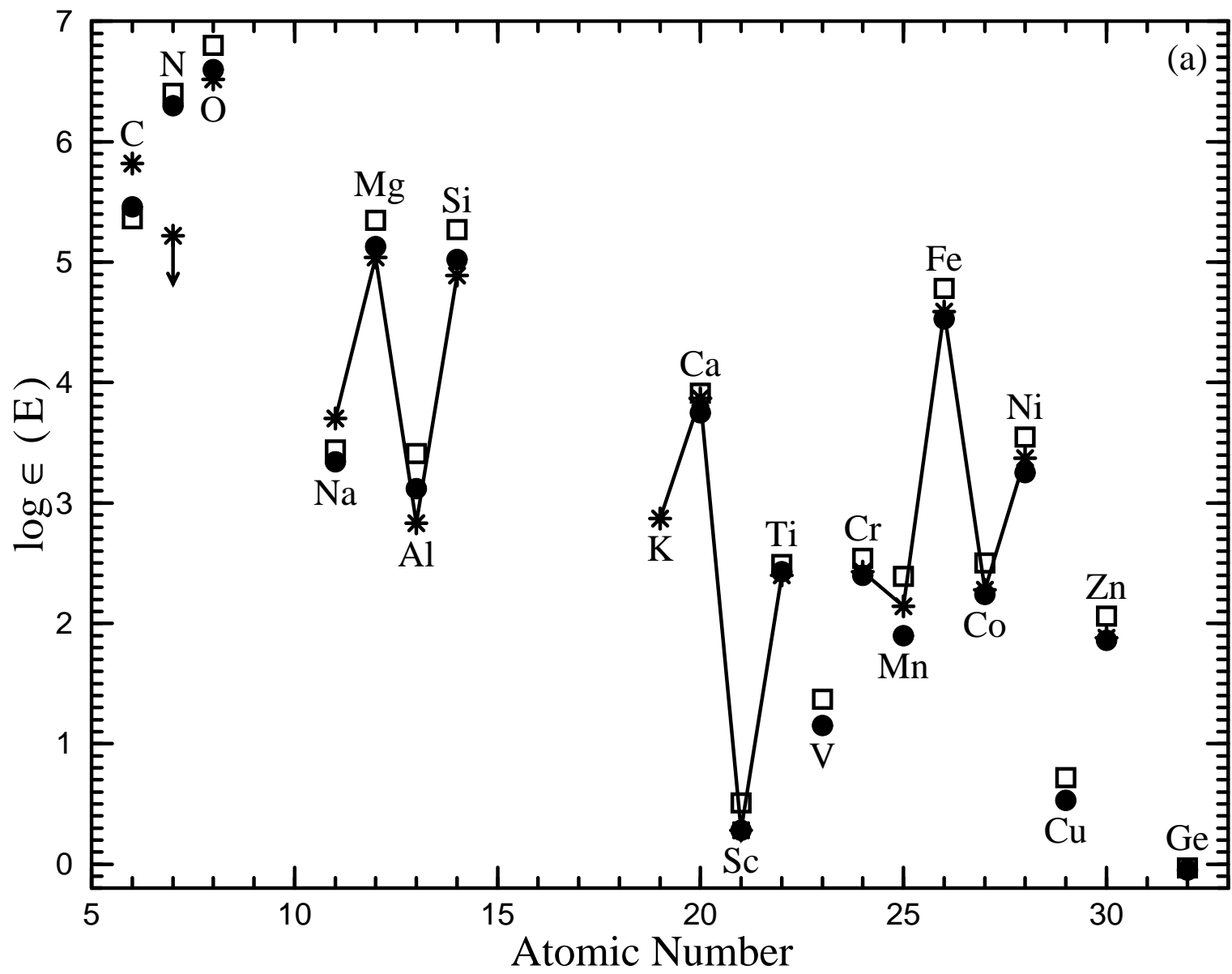
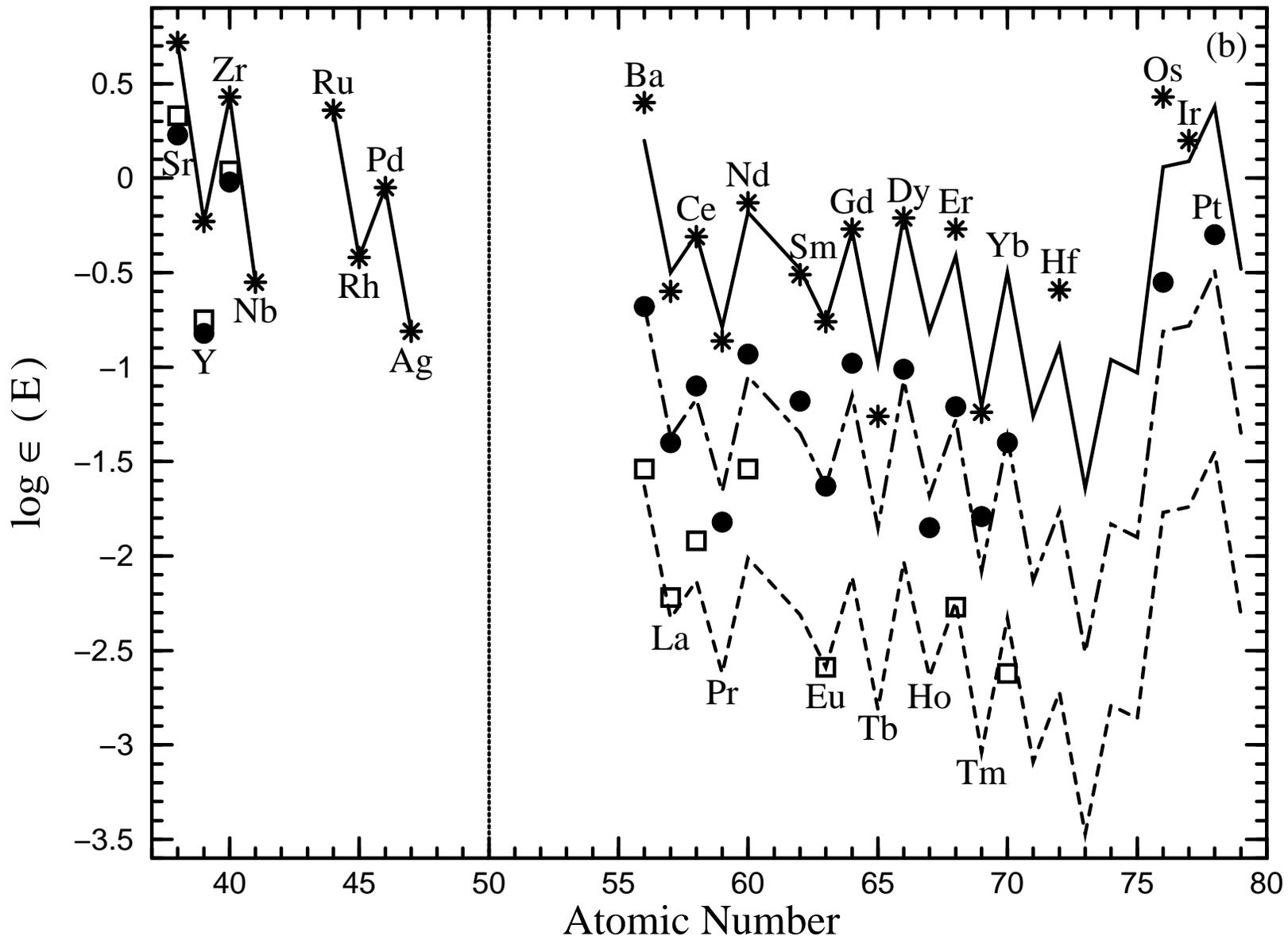
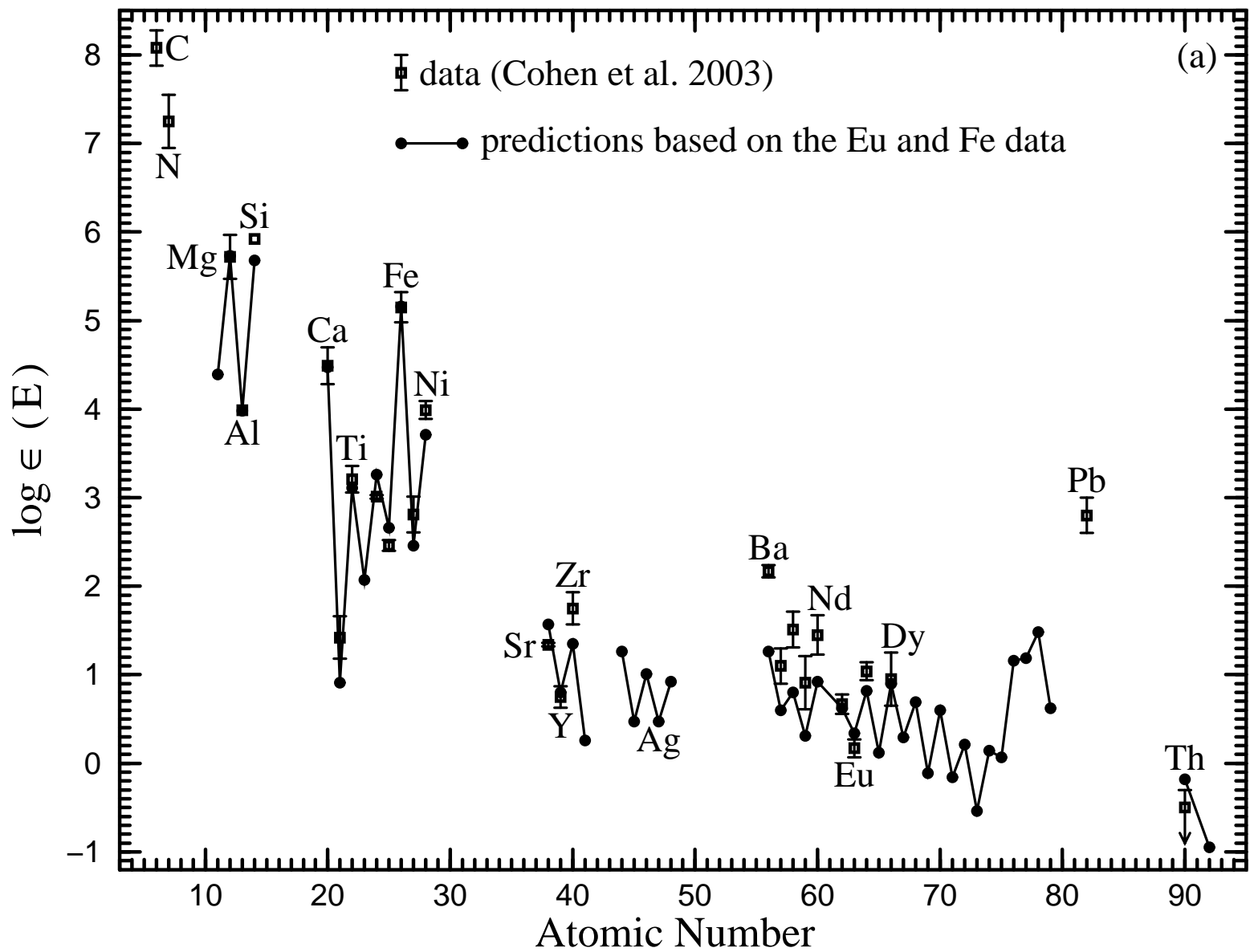
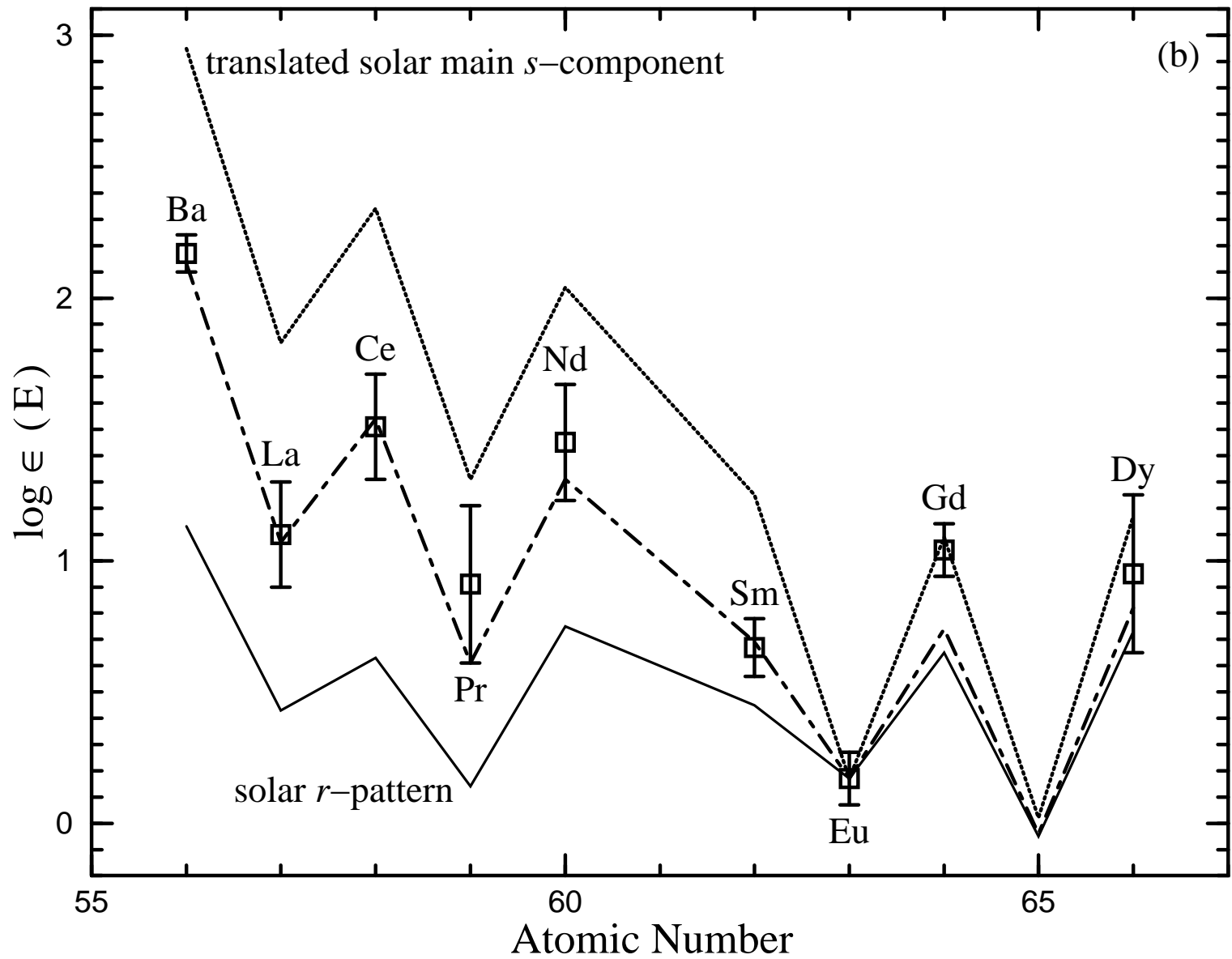
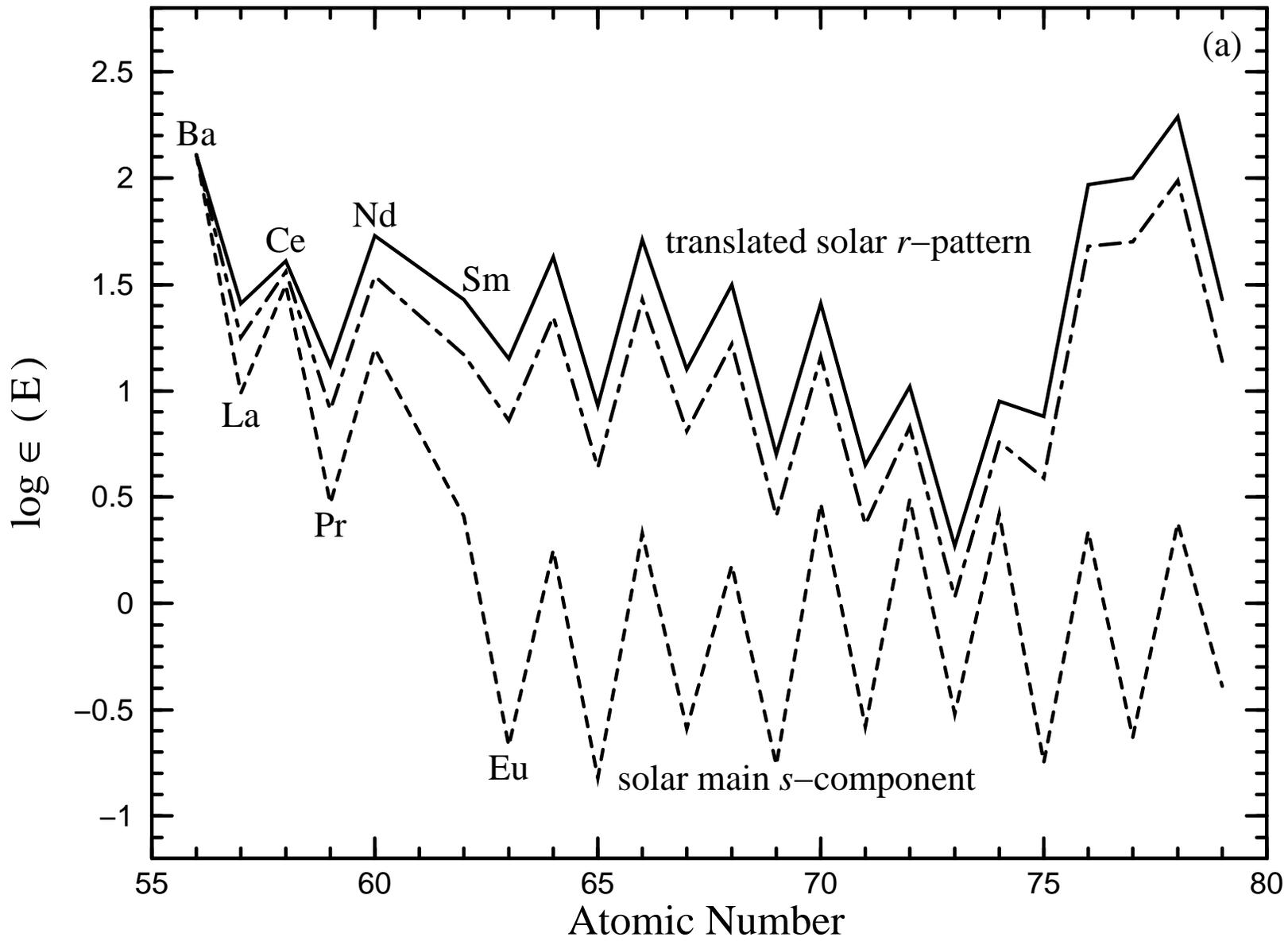
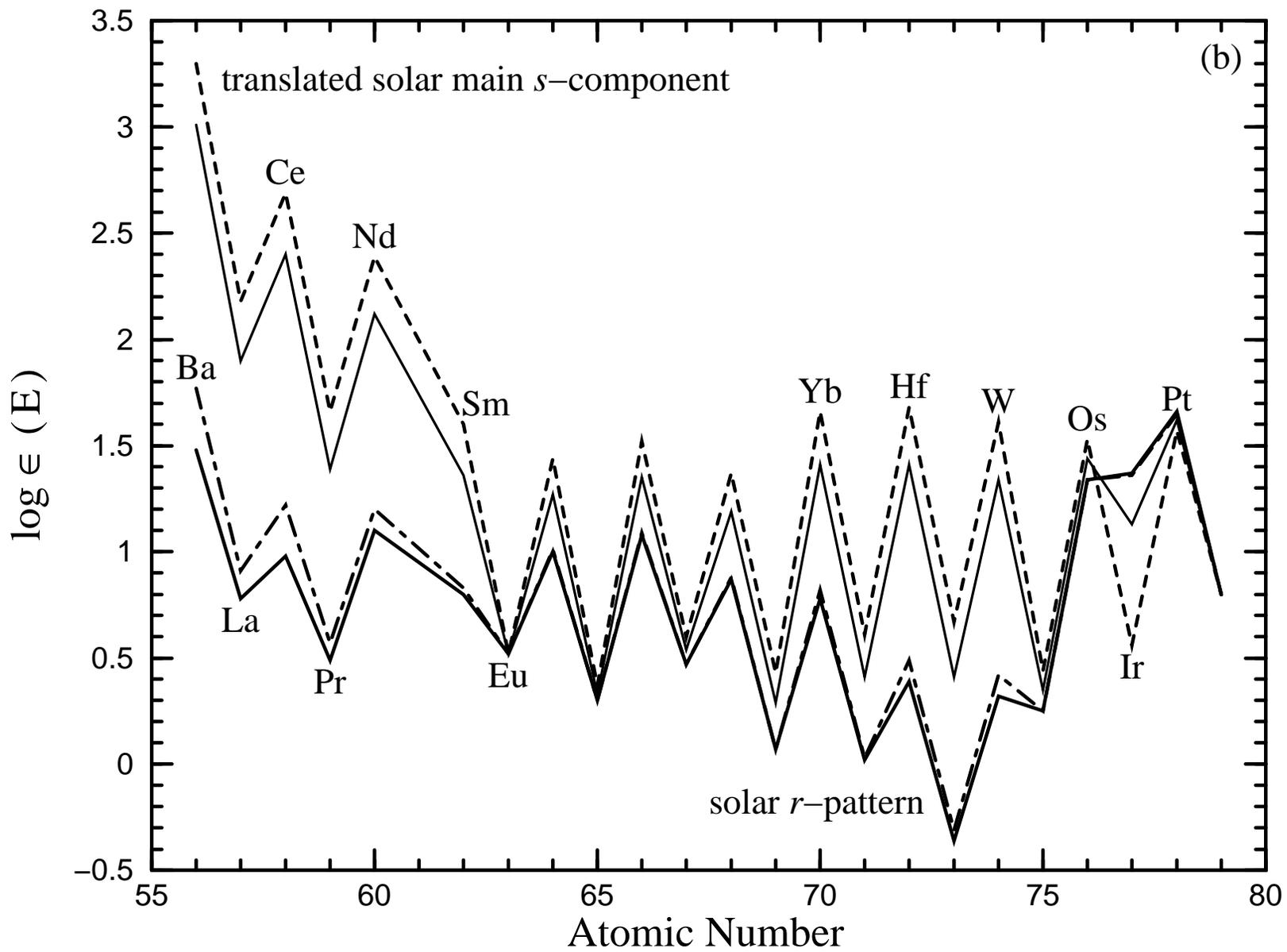
\end{document}